\documentclass[11pt,twoside]{article}
\usepackage{asp2006}
\usepackage{graphics}
\usepackage{lscape}
\def\arg#1{{\it#1\/}}
\let\prog=\arg
\def\edcomment#1{\iffalse\marginpar{\raggedright\sl#1\/}\else\relax\fi}
\reversemarginpar

\begin{document}
\title{Where can we find Super-Earths?}
\author{E. Podlewska-Gaca, E. Szuszkiewicz}
\affil{CASA* and Institute of Physics, University of Szczecin, Poland}

\begin{abstract}
  In recent years we have been witnessing the discovery of one extrasolar
  gas giant after another. Now the time has come to detect more low-mass
  planets like Super-Earths and Earth-like objects. An  interesting
  question to ask is: where should we look for them? We have explored here
  the possibility of finding Super-Earths in the close vicinity of gas
  giants, as a result of the early evolution  of  planetary systems.
  For this purpose, we have considered a young planetary system containing
  a Super-Earth and a gas giant, both embedded in a protoplanetary disc.
  We have shown that, if the Super-Earth is on the internal orbit relative to
  the 
  gas giant, 
  the planets can easily become locked in a mean motion resonance. This is
  no longer true, however, if the Super-Earth is on the  external orbit. In
  this case we have obtained that the low-mass planet is captured in a trap
  at the outer edge 
  of the gap opened by the giant planet and no first  order mean motion 
  commensurabilities are expected.
  Our investigations might be particularly
  useful for the observational TTV (Transit Timing Variation) technique.
\end{abstract}

The population of Super-Earths (planets with the  mass in the range of 2-10
$M_{\oplus}$) should be numerous according to theories of
planet formation (Ida \& Lin, 2005). Till now we know 18 such objects around main sequence
stars. In comparison, there are more than 300 gas giant planets in the 
solar neighbourhood. So, where are the Super-Earths? 
Most of the known low-mass planets have been discovered not later than 2 years
ago, thanks to the sensitive instruments and sophisticated observational
techniques used from the ground (e.g. HARPS at La Silla Observatory, 
microlensing
programmes OGLE, MOA) and from the space (CoRoT). Another
space mission, Kepler, will soon help to improve the statistics.
Following the successful hunting for Super-Earths, we have undertaken
the task of answering the question whether there are any preferred 
planetary configurations in which Super-Earths are involved.
We have started our investigation from a particular
configuration, namely from a Super-Earth orbiting a Sun-like star close
to a gas giant. We have given most emphasis to the possible resonant 
structures
in such system.

In our studies we have performed a series of 2D hydrodynamic simulations 
of a pair of 
interacting planets: a Super-Earth and a Jupiter, both embedded in a gaseous
 protoplanetary disc.
The migration process induced by the disc-planet interactions will
determine 
the final architecture of this system. Particularly interesting for us
here, is  the convergent
migration, which  is one of the most promising mechanisms for bringing
planets into 
mean motion resonances.

The outcome of our simulations is illustrated in Fig. 1. The Jupiter is located at the
distance 
1 (in dimensionless units) from the star, which corresponds to 5.2 AU in 
the Solar System.
The results suggest that we can find Super-Earths either close to the 
locations of the
inner first order mean motion resonances 
with a gas giant 
(Podlewska \& Szuszkiewicz, 2008)  
or at the position 
of the external edge of the gap opened
by the Jupiter like planet  (Podlewska \& Szuszkiewicz, 2009). 
The reason for the latter case is that the Super-Earth
is caught in a trap at the  edge of the gap.
A similar planet behaviour has been found by  Masset et al. (2006), Pierens \&
Nelson (2008) 
and more recently it has been explored in detail by
Paardekooper \& Papaloizou (2009). 
Our investigations 
might
be particularly  useful for the observational TTV technique (Agol et al., 
2005) which is
sufficiently powerful to detect the planets as small as our Earth,
especially if it is locked in the mean motion resonance with the gas giant.

\begin{figure}
\centering
\vskip 8.0cm
\includegraphics{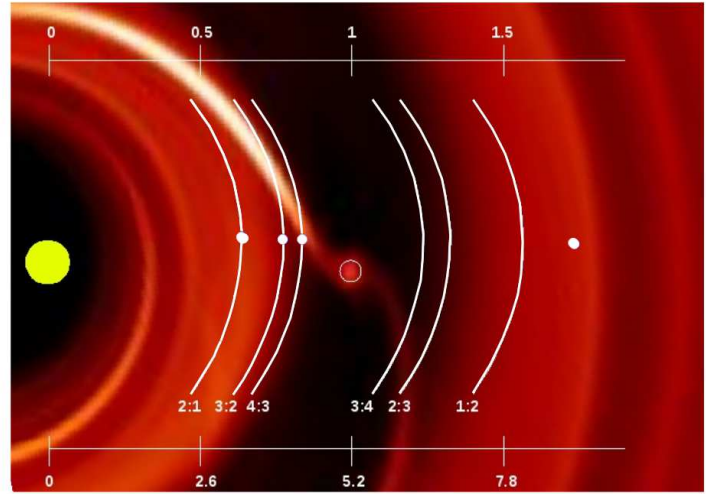}
\centerline{
}
\caption{\label{resonances1}{
The possible locations of the Super-Earth (denoted as white dots) in the 
system containing the
Jupiter mass gas giant.
}}
\end{figure}

\begin{acknowledgments}
This work has been partially supported by grants: MNiSW (N203 026 32/3831) 
and ASTROSIM-PL. The simulations reported here were performed using the 
HPC cluster HAL9000 of the Computing Centre of the Faculty of Mathematics
and Physics at the University of Szczecin.   
\end{acknowledgments}

\vfill\pagebreak

\end{document}